\documentstyle[preprint,aps]{revtex}
\tightenlines
\begin{document}
\title{Transfer matrix of a spherical scatterer}
\author{V. S. Podolsky and A. A. Lisyansky}
\address{Department of Physics, Queens\\
College of City University of New York, Flushing, New York 11367}
\date{\today}
\maketitle

\begin{abstract}
We derive the off-shell scattering matrix for a spherical scatterer. The
result obtained generalizes the off-on-shell matrix commonly used in the
theory of scalar waves propagation in random media.
\end{abstract}


\section{introduction}

It was recently recognized that the internal structure of scatterers can
significantly alter the picture of wave propagation and localization in
random media \cite{Kirpatrick,Albada}. When the size of scatterers cannot be
neglected, scattering is anisotropic and a ``dwell time'' for waves appears
in addition to the time of flight between scattering \cite{Albada,Wigner}.
This substantially modifies such parameters describing wave propagation as
the diffusion coefficient \cite
{Albada,Boer,Bart,Kogan,Cwilich,Kroha,Soukoulis,Liv1,Liv2}, transport mean
free path \cite{Liv1,Liv2} and transport velocity \cite
{Ozrin,Comment,Rigorous}. The microscopic approach to the problem of the
renormalization of the diffusion coefficient is based on the Bethe-Salpeter
equation for the field-field correlation function. This equation and the
solution obtained involve the transfer matrix for a single scatterer, ${%
\mbox{\Large $t$}}^{\omega}_{{\bf k\,k^{\prime}}}$, where ${\bf k,k^{\prime}}
$ are outgoing and incident momenta, respectively, and $\omega$ is the
frequency \cite{Sheng}. This matrix contains all information relating to a
scatterer's structure, its influence on the anisotropy of scattering and on
internal resonances. A great deal of this information is ignored in the
commonly used {\em on-shell\/} approximation for the transfer matrix where
both momenta ${\bf k}$ and ${\bf k^{\prime}}$ are taken on the ``mass
shell'', $k^{2}\!=\!{\omega}^{2}$. The {\em off-shell\/} transfer matrix,
originally arising in the Bethe-Salpeter equation, is not restricted by this
condition. This matrix resolves the structure of a scattered field at any
distance away from a scatterer, while the on-shell matrix gives the far-zone
asymptote of a field only. The on-shell approximation formally enters the
theory of wave transport in random media via the commonly accepted $\delta$%
-function approximation for the imaginary part of the Green function, ${\rm %
Im}G_{k,\omega}\propto\delta(\omega^{2}\!\!-\!c^{2}_{p}k^{2})$. The $\delta$
approximation confines all transfer matrices in the expression of the
diffusion coefficient on the mass shell. However, the presence of
derivatives of ${\mbox{\Large
$t$}}^{\omega}_{{\bf k\,k^{\prime}}}$ does not allow one to set $\omega=k$
automatically. Even within the on-shell approximation, one has to first
calculate the derivatives and then to allow $k\rightarrow\omega$. Neglecting
this fact can lead to unexpected results. For instance, numerical
calculations of the influence of weak absorption within the on-shell
approximation show that the diffusion coefficient becomes larger than its
value in the medium without microstructure \cite{Liv95}. This effect
disappears if the off-shell matrix is used \cite{Liv1,Liv2}.

The {\em off-on-shell\/} matrix for a scatterer with an infinite
permittivity was first used in Ref. \cite{Kirpatrick}. Such a matrix gives a
far-zone asymptote of a scattered field when the source is located at a
finite distance from a scatterer. This destroys the symmetry between the
incident and scattered momenta which exists due to the reciprocal principal
for the Green function. Analysis of the renormalization of the diffusion
coefficient employing the off-on-shell transfer matrix of a scatterer with a
finite permitivity \cite{Liv1,Liv2} has shown a strong enhancement of the
previously obtained \cite{Albada,Cwilich} corrections to $D$ and
qualitatively agrees with experimental data \cite{Genack}. Numerical
agreement can be attained with the use of an exact off-shell matrix. In this
paper we derive the transfer matrix of a dielectric sphere and analyze some
of its general properties.

\section{transfer matrix of a dielectric sphere}

The transfer matrix of a single scatterer is originally introduced via the
Green function of the wave equation 
\begin{equation}
\left\{{\omega}^2\left[1+\varepsilon({\bf r)}\right]+\triangle\right\}G^{}_{%
{\bf r,r^{\prime}}}=-\delta^{}_{{\bf r-r^{\prime}}},
\end{equation}
where ${\bf r,r^{\prime}}$ are location vectors of the point of observation
and the source, respectively; $\varepsilon(r)$ has a constant value, $%
\varepsilon$, inside the scatterer and vanishes outside, the wave speed in
this equation and below is equal to unity. Eq.\ (1) must be solved under
conditions that $G^{}_{{\bf r,r^{\prime}}}$ and its derivatives $\partial
G^{}_{{\bf r,r^{\prime}}}/\partial{\bf r}$ are continuous in the entire
space outside the source \cite{Morse}. For a spherical scatterer, one can
obtain $G^{}_{{\bf r,r^{\prime}}}$ by utilizing separation of the spherical
coordinates and using proper boundary conditions at the surface of the
scatterer 
\begin{eqnarray}
G^{}_{{\bf r,r^{\prime}}}={\frac{{\Large e}_{}^{i\omega R}}{ 4\pi R}
+i/4\sum^{\infty}_{l=0}\frac{(l+1/2)P({\bf \hat{r}\hat{r}^{\prime}}%
)H^{}_{\omega r} H^{}_{\omega r^{\prime}} }{\sqrt{rr^{\prime}}}%
A^{}_{l}\:\:\:\: {\rm when}\:\: r,r^{\prime}\ge a, }  \nonumber \\
G^{}_{{\bf r,r^{\prime}}}={i/4\sum^{\infty}_{l=0}\frac{(l+1/2)P({\bf \hat{r}%
\hat{r}^{\prime}})J^{}_{\Omega r} H^{}_{\omega r^{\prime}} }{\sqrt{%
rr^{\prime}}}B^{}_{l}\:\: \:\: \:\: {\rm when}\:\: r\le a,r^{\prime}\ge a, }
\nonumber \\
{} \\
G^{}_{{\bf r,r^{\prime}}}={i/4\sum^{\infty}_{l=0}\frac{(l+1/2)P({\bf \hat{r}%
\hat{r}^{\prime}})H^{}_{\omega r} J^{}_{\Omega r^{\prime}} }{\sqrt{%
rr^{\prime}}}B^{}_{l}\:\: \:\: \:\: {\rm when}\:\: r\ge a,r^{\prime}\le a, }
\nonumber \\
G^{}_{{\bf r,r^{\prime}}}={\frac{{\Large e}_{}^{i\Omega R}}{ 4\pi R}
+i/4\sum^{\infty}_{l=0}\frac{(l+1/2)P({\bf \hat{r}\hat{r}^{\prime}}%
)J^{}_{\Omega r} J^{}_{\Omega r^{\prime}} }{\sqrt{rr^{\prime}}}%
C^{}_{l}\:\:\:\: {\rm when}\:\: r,r^{\prime}\le a, }  \nonumber
\end{eqnarray}
where $a$ is the radius of the scatterer, $R\!=\!|{\bf r\!-\!r^{\prime}}|$
is the distance between the point of observation and the source and ${\Omega}%
^{2}\!=\!{\omega}^{2}(1\!+\!\varepsilon)$. In these equations and below we
omit the index {\em l} denoting $P(x)\!=\!P^{}_{l}(x)$, $J^{}_{x}\!=J^{}_{l%
\!+\!1/2}(x)$ and $H^{}_{x}\!=\!H^{\scriptscriptstyle+}_{l\!+\!1/2}(x)$,
where $P^{}_{l}(x)$ are the Legendre polynomials, $J^{}_{l\!+\!1/2}(x)$ are
the Bessel functions and $H^{\scriptscriptstyle+}_{l\!+\!1/2}(x)$ are the
Hankel functions of the first kind. The coefficients of these expansions,
obtained by boundary matching, have the form: 
\begin{equation}
A^{}_{l}\!=\!-\frac{|J^{}_{\omega a}J^{}_{\Omega a}|}{|H^{}_{\omega
a}J^{}_{\Omega a}|},\:\: C^{}_{l}\!=\!-\frac{|H^{}_{\omega a}H^{}_{\Omega a}|%
}{|H^{}_{\omega a}J^{}_{\Omega a}|},\:\: B^{}_{l}\!=\!-\frac{2i/\pi a}{%
|H^{}_{\omega a}J^{}_{\Omega a}|}.
\end{equation}
Here we use a short notation of the following determinant based on any pair
of cylindrical functions $U^{}_{\nu}(x)$ and $V^{}_{\nu}(x)$ 
\begin{equation}
|U^{}_{\alpha x}V^{}_{\beta x}|=\left|
\begin{array}{cc}
{U^{}_{\nu}(\alpha x)} & {\ V^{}_{\nu}(\beta x)} \\ 
{\alpha U^{}_{\nu+1}(\alpha x)} & {\beta V^{}_{\nu+1}(\beta x)}
\end{array}
\right|
\end{equation}
By definition, the transfer matrix and the Green function are related by the
equation, 
\mbox{$G^{}_{\bf k\,k'}\!=\!\delta^{}_{\bf k k'}D^{\omega}_{k}+{\mbox{\Large
$t$}}^{\omega}_{\bf k\,k'}D^{\omega}_{k}D^{\omega}_{k'}$}, where $%
D^{\omega}_{k}\!=\left[k^{2}\!-{\omega}^{2}\right]^{-1}$ is the wave
propagator in empty space. Calculating the Fourier transform of the Green
function, \ $\int^{}_{{\bf r}}\int^{}_{{\bf r^{\prime}}}\left(2\pi%
\right)^{-3}G^{}_{{\bf r,r^{\prime}}}\exp{(-i{\bf kr}\!+\!i{\bf %
k^{\prime}r^{\prime}})}$, \ and using expansions of the spherical and plane
waves over the spherical functions, $Y^{}_{lm}$, one can obtain 
\begin{eqnarray}
{{\mbox{\Large $t$}}^{\omega}_{{\bf k\,k^{\prime}}}D^{\omega}_{k}\!D^{%
\omega}_{k^{\prime}}= \sum^{\infty}_{l=0}\!\frac{i(l\!+\!1/2)P({\bf \hat{k}%
\hat{k}^{\prime}})}{4\sqrt{kk^{\prime}}}\!
\left[\int^{\infty}_{0}\!\left(D^{\Omega}_{q}\!-\!D^{\omega}_{q}\right)\frac{%
dq^2}{i\pi}\!
\int^{a}_{0}\!rJ^{}_{kr}J^{}_{qr}dr\!\int^{a}_{0}\!r^{\prime}J^{}_{k^{%
\prime}r^{\prime}}J^{}_{qr^{\prime}}dr^{\prime}\right.}  \nonumber \\
{}  \nonumber \\
{+\int^{\infty}_{a}\!rJ^{}_{kr}H^{}_{\omega
r}dr\!\int^{a}_{0}\!r^{\prime}J^{}_{k^{\prime}r^{\prime}}J^{}_{\omega
r^{\prime}}dr^{\prime}+\int^{a}_{0}\!rJ^{}_{kr}J^{}_{\omega
r}dr\!\int^{\infty}_{a}\!r^{\prime}J^{}_{k^{\prime}r^{\prime}}H^{}_{\omega
r^{\prime}}dr^{\prime}} \\
{}  \nonumber \\
{\left. +\left( \!\int^{\infty}_{a}\!rJ^{}_{kr}H^{}_{\omega
r}dr;\int^{a}_{0}\!rJ^{}_{kr}J^{}_{\Omega r}dr\! \right)\! \left(\!
\begin{array}{cc}
A^{}_{l} & B^{}_{l} \\ 
B^{}_{l} & C^{}_{l}
\end{array}
\!\right)\!\left(\! 
\begin{array}{c}
{\int^{\infty}_{a}\!r^{\prime}J^{}_{k^{\prime}r^{\prime}}H^{}_{\omega
r^{\prime}}dr^{\prime}} \\ 
{\int^{a}_{0}\!r^{\prime}J^{}_{k^{\prime}r^{\prime}}J^{}_{\Omega
r^{\prime}}dr^{\prime}}
\end{array}
\!\right)\right]}  \nonumber
\end{eqnarray}
The integrals of the Bessel functions arising in Eq.\ (5) can be found in
Ref.\ \cite{Abramowitz} 
\begin{equation}
\int^{a}_{0}xJ^{}_{\alpha x}J^{}_{\beta x}dx=\frac{a}{\beta_{}^{2}\!-\!%
\alpha_{}^{2}}\left|J^{}_{\alpha x}J^{}_{\beta x}\right|
\end{equation}
The integrals containing Hankel functions diverge on the upper limit.
However, since we deal with the retarded Green function, the frequency is
shifted in the upper half-plane, $\omega\!\rightarrow\!\omega\!+\!i0$. This
gives an exponentially decaying factor in $H^{}_{\omega r}\:{\rm when}%
\:r\!\rightarrow\!\infty$ and provides convergence of the integrals. Using
recurrence relations between cylindrical functions one can show 
\begin{equation}
\int^{\infty}_{a}xJ^{}_{\alpha x}H^{}_{\beta x}dx=\frac{a}{%
\alpha_{}^{2}\!-\!\beta_{}^{2}}\!\left|J^{}_{\alpha x}H^{}_{\beta x}\right|
\end{equation}
Using Eqs.\ (3), (6), and (7) one can represent the {\em off-shell\/}
transfer matrix of a dielectric sphere in the form 
\begin{eqnarray}
{{\mbox{\large$t$}}^{\omega}_{{\bf k\,k^{\prime}}}= \sum^{\infty}_{l=0}\!%
\frac{ia^{2}(l\!+\!1/2)\!P({\bf \hat{k}\hat{k}^{\prime}})}{4\sqrt{kk^{\prime}%
}}\! \left[L^{\omega}_{k}L^{\omega}_{k^{\prime}}\!\int^{\infty}_{0}\! \frac{%
i\varepsilon{\omega}^{2}dq^{2}\!\left|J^{}_{ka}J^{}_{qa}\right|
\left|J^{}_{k^{\prime}a}J^{}_{qa}\right|} {\pi\left(q^{2}\!-\!k^{2}\right)\!
\left(q^{2}\!-\!{k^{\prime}}^{2}\right)\! \left(q^{2}\!-\!{\omega}%
^{2}\right)\! \left(q^{2}\!-\!{\Omega}^{2}\right)}\right.}  \nonumber \\
{}  \nonumber \\
{+\left|J^{}_{ka}H^{}_{\omega a}\right|\left|J^{}_{k^{\prime}a}J^{}_{\omega
a}\right|+ \left|J^{}_{ka}J^{}_{\omega
a}\right|\left|J^{}_{k^{\prime}a}H^{}_{\omega a}\right|} \\
{}  \nonumber \\
{\left.-\frac{1}{\left|H^{}_{\omega a}J^{}_{\Omega a}\right|}
\left(-\!\left|J^{}_{ka}H^{}_{\omega a}\right|;
L^{\omega}_{k}D^{\Omega}_{k}\left|J^{}_{ka}J^{}_{\Omega a}\right|\right)
\!\left(
\begin{array}{cc}
{\left|J^{}_{\omega a}J^{}_{\Omega a}\right|} & {2i/\pi a}\nonumber \\ 
{2i/\pi a} & {\left|H^{}_{\omega a}H^{}_{\Omega a}\right|}\nonumber
\end{array}
\!\!\right)\!\!\left(\!\! 
\begin{array}{c}
{-\,\left|J^{}_{k^{\prime}a}H^{}_{\omega a}\right|}\nonumber \\ 
{L^{\omega}_{k^{\prime}}D^{\Omega}_{k^{\prime}}\left|J^{}_{k^{%
\prime}a}J^{}_{\Omega a}\right|}\nonumber
\end{array}
\!\!\right)\!\right],}  \nonumber
\end{eqnarray}
where $L^{\omega}_{k}\!=\!k^{2}\!-\!{\omega}^{2}$ is the inverse wave
propagator in empty space. This expression has an explicit symmetry, ${%
\mbox{\large$t$}}^{}_{{\bf k\,k^{\prime}}} \!=\!{\mbox{\large$t$}}^{}_{{\bf %
k^{\prime}\,k}}$, as it must for a spherical scatterer. Eq.\ (8) cannot be
simplified any more in a general situation.

The {\em off-on-shell\/} matrix can be obtained from Eq.\ (8) by taking the
outgoing momentum to the mass shell, ${\bf k}\!=\!\omega{\bf \hat{k}}$, and
using the facts that $\left|J^{}_{\omega a}J^{}_{\omega
a}\right|\!=\!L^{\omega}_{\omega}=\!0$ and $\left|J^{}_{\omega
a}H^{}_{\omega a}\right|\!=-2i/\pi a$ 
\begin{equation}
{\mbox{\large$t$}}^{\omega}_{\omega{\bf \hat{k}\,k^{\prime}}%
}\!=\!\sum^{\infty}_{l=0} \!\frac{a(l\!+\!1/2)\!P({\bf \hat{k}\hat{k}%
^{\prime}})}{2\pi\sqrt{\omega k^{\prime}}}\!\left[
\left|J^{}_{k^{\prime}a}J^{}_{\omega a}\right|+ \frac{1}{\left|H^{}_{\omega
a}J^{}_{\Omega a}\right|}\! \left(\left|J^{}_{\omega a}J^{}_{\Omega
a}\right|;2i/\pi a\right) \left(\!
\begin{array}{c}
{-\left|J^{}_{k^{\prime}a}H^{}_{\omega a}\right|} \\ 
{L^{\omega}_{k^{\prime}}D^{\Omega}_{k^{\prime}}\left|J^{}_{k^{%
\prime}a}J^{}_{\Omega a}\right|}
\end{array}
\!\right)\!\right]
\end{equation}
The transfer matrix given by Eq.\ (9) coincides with the result of Refs. 
\cite{Liv1,Liv2}. In the limit $\varepsilon\rightarrow\infty$ ($%
\Omega\rightarrow\infty$) it reproduces the result of Ref. \cite{Kirpatrick}%
. This transfer matrix defines the far-zone asymptote of the scattered field
in the case that the point source is located at a finite distance from the
scatterer. That destroys the symmetry between the incident and scattered
momenta.

For a source at an infinite distance we have to take the incident momentum, $%
{\bf k^{\prime}}$, in Eq.\ (9) to the mass shell. This results in the
well-known expression for the {\em on-shell\/} transfer matrix of a
spherical dielectric scatterer \cite{Van de Hulst} 
\begin{equation}
{\mbox{\large$t$}}^{\omega}_{{\bf \hat{k}\hat{k}^{\prime}}%
}=\sum^{\infty}_{l=0} \frac{i(l\!+1/2)P({\bf \hat{k}\hat{k}^{\prime}}%
)\left|J^{}_{\omega a}J^{}_{\Omega a}\right|} {\pi^2\omega\left|H^{}_{\omega
a}J^{}_{\Omega a}\right|}.
\end{equation}

\section{Rayleigh limit}

All expressions given by Eqs.\ (8-10) can be substantially simplified in the
Rayleigh limit when the scatterer size is the smallest linear scale in a
problem. Taking into account conditions $\omega a,\,\Omega
a,\,ka,\,k^{\prime}a\ll 1$ and using the asymptotes for cylindrical
functions one can show that 
\begin{eqnarray*}
{\left|\!J^{}_{\alpha a}J^{}_{\beta a}\right|\! \approx\!\!\frac{%
a\!\left(\beta^{2}\!\!-\!\alpha^{2}\right)\!\left(\!\frac{\alpha\beta a^{2}}{%
4}\right)^{l+1/2}\!}{2\Gamma(l\!+\!3/2)\Gamma(l\!+\!5/3)},\:\:
\left|H^{}_{\alpha a}J^{}_{\beta a}\right|\!\approx\! \frac{-2}{ai\pi}%
\!\!\left(\!\frac{\beta}{\alpha}\!\right)^{l+1/2},} \\
{} \\
{\left|H^{}_{\alpha a}H^{}_{\beta a}\right|\!\approx\! \frac{%
a\!\left(\beta^2\!\!-\!\alpha^2\right)\!\Gamma^{2}(l\!+\!1/2)\left(\!\frac{%
\alpha\beta a^{2}}{4}\!\right)^{-(l+1/2)}}{2\pi^2\!\left(l\!-\!1/2\right)}\:.%
}
\end{eqnarray*}
Substitution of these equations into Eq.\ (8) yields the transfer matrix for
long waves 
\begin{eqnarray}
{{\mbox{\large$t$}}^{\omega}_{{\bf k\,k^{\prime}}}\approx \sum^{\infty}_{l=0}%
\frac{a^2(l\!+\!1/2) P({\bf \hat{k}\hat{k^{\prime}}})\left(\frac{%
kk^{\prime}a^2}{4}\right)^{l+1/2}}{4\pi\sqrt{kk^{\prime}}\Gamma(l\!+\!3/2)%
\Gamma(l\!+\!5/2)}\!\left[2\left(L^{\omega}_{k}\!+\!L^{\omega}_{k^{\prime}}%
\right)\!+\!\varepsilon{\omega}^2\left(\!1\!+\! \frac{a^{2}L^{\omega}_{k}L^{%
\omega}_{k^{\prime}}}{4\left(l+1/2\right)^2\left[\left(l+1/2\right)^2\!-\!1%
\right]}\right)\right.}  \nonumber \\
{} \\
{\left.\!\!+\varepsilon{\omega}^2L^{\omega}_{k}L^{\omega}_{k^{\prime}}\!%
\int^{\infty}_{0} \frac{\frac{q^2\left(2l+3\right)}{2}%
\!J^{{}^{2}}_{l+3/2}(qa)\!-\frac{qa\left(k^2\!+{k'}^2\right)}{2}%
J^{}_{l+1/2}(qa)J^{}_{l+3/2}(qa)\!+ \frac{\left(kk^{\prime}a^{\prime}%
\right)^{2}}{2\left(2l+3\right)}J^{{}^{2}}_{l+1/2}(qa)} {\left(q^2\!-\!k^2%
\right)\left(q^2\!-\!k^{'2}\right)
\left(q^2\!-\!\omega^2\right)\left(q^{2}\!-\!\Omega^2\right)}dq^2\right].} 
\nonumber
\end{eqnarray}
Due to the presence of two factorial functions in the denominator of Eq.\
(11), the main term in the sum is the term with $l=0$. This shows that in
the Rayleigh limit an isotropic $S$ term dominates in the scattered field
even at distances of the order of the wavelength from the scatterer.
Calculating the isotropic term of the transfer matrix at the mass shell we
reproduce the well-known Rayleigh scattering amplitude, $t(\omega)\!=%
\varepsilon\omega^2 a^3/{6\pi^2}$.

Eq.\ (11) contains three types of terms with different behavior at the mass
shell. The first type of term remains finite when $k,k^{\prime}\!\rightarrow%
\omega$, the second type vanishes as $L^{\omega}_{k}$, and the third type
vanishes as $\left({L^{\omega}_{k}}\right)^2$. The third type of term were
usually missed in calculations, since the off-on-shell transfer matrix was
used instead of the off-shell matrix. This leads to incorrect values of the
partial derivatives $\partial{\mbox{\Large $t$}}^{\omega}_{{\bf k k}%
}/\partial\omega^{2}|_{k=\omega}\:\:{\rm and}\:\:\partial{\mbox{\Large $t$}}%
^{\omega}_{{\bf k k}}/\partial k^{2}|_{k=\omega}$ in the expression for the
diffusion coefficient. Using Eqs.\ (8), (11) enables one to obtain correct
results for these derivatives. Below we obtain some general relations for $%
\partial{\mbox{\Large $t$}}^{\omega}_{{\bf kk}}/\partial\omega^{2}$ and $%
\partial{\mbox{\Large $t$}}^{\omega}_{{\bf kk}}/\partial {\bf k}$ which
allow one to avoid lengthy straightforward calculations.

\section{Derivatives of the t matrix}

Calculating the derivative of the operator equation $G\!=\!D\!+\!D{%
\mbox{\large$t$}}D$ with respect to $\omega^{2}$ and using the fact that $%
\partial D/\partial\omega^{2}\!=\! D^{2}$ one can obtain 
\begin{equation}
\frac{\partial G}{\partial\omega^{2}}\!=\!D^{2}\!+\!D^{2}{\mbox{\large$t$}}%
D\!+\!D{\mbox{\large$t$}}D^{2}\!+\!D\frac{\partial {\mbox{\large$t$}}}{%
\partial\omega^{2}}D
\end{equation}
On the other hand, differentiation of Eq.\ (1) yields 
\begin{equation}
\frac{\partial G}{\partial\omega^{2}}\!=\!G\left(\!1\!+{\cal E}%
\right)G\!=\!\left(D\!+\!D{\mbox{\large$t$}}D\right)\left(1\!+\!{\cal E}%
\right)\!\left(D\!+\!D{\mbox{\large$t$}}D\right)
\end{equation}
Here the operators $D\:\:{\rm and}\:\:{\cal E}$ are diagonal in the momentum
and coordinate representations, respectively, $\langle{\bf k}|D|{\bf %
k^{\prime}}\rangle\!=\!\delta_{{\bf k\,k^{\prime}}}\left(k^{2}\!-\!%
\omega^{2}\right)^{-1},$ $\langle{\bf x}|{\cal E} | {\bf x^{\prime}}%
\rangle\!=\!\varepsilon({\bf x})\delta_{{\bf x\,x^{\prime}}}$. From the
perturbation series expansion of the scattering operator, ${\mbox{\large$t$}}%
=\! \omega^{2}{\cal E}\! +\omega^{4}{\cal E} D{\cal E}\!+\ldots\,,$ it
follows that ${\cal E}D{\mbox{\large$t$}}\!=\!{\mbox{\large$t$}}D{\cal E}%
\!=\!{{\mbox{\large$t$}}/{\omega^2}-\!{\cal E}}$. Using this relation to
eliminate ${\cal E}$ from Eq.\ (13) and comparing the result with Eq.\ (12),
we obtain the operator equation 
\begin{equation}
\frac{\partial {\mbox{\large$t$}}}{\partial\omega^{2}}\!=\!\frac{{%
\mbox{\large$t$}}}{\omega^2}\!+\!{\mbox{\large$t$}}\left(D^{2}\!+\!\frac{D}{%
\omega^2}\right){\mbox{\large$t$}}.
\end{equation}
Using the explicit form for matrix elements of all operators in Eq.\ (14),
we obtain an expression for the derivative $\partial{\mbox{\Large $t$}}%
^{\omega}_{{\bf k k^{\prime}}}/\partial\omega^{2}$ 
\begin{equation}
\frac{\partial {\mbox{\large$t$}}^{\omega}_{{\bf k k^{\prime}}}}{%
\partial\omega^{2}}\!= \!\frac{{\mbox{\large$t$}}^{\omega}_{{\bf k k^{\prime}%
}}}{\omega^2}\!+ \!\int{\mbox{\large$t$}}^{\omega}_{{\bf k q}}{%
\mbox{\large$t$}}^{\omega}_{{\bf q k^{\prime}}}\frac{q^{2}d{\bf q}}{%
\omega^{2}\left(q^{2}\!-\!\omega^{2}\right)^{2}}.
\end{equation}
A similar formula for derivatives of ${\mbox{\Large $t$}}^{\omega}_{{\bf k
k^{\prime}}}$ with respect to wave vectors or their combinations can be
obtained by differentiating the momentum representation for the perturbation
expansion of the scattering operator, 
\[
{\mbox{\large$t$}}^{\omega}_{{\bf k k^{\prime}}}=\! \omega^{2}{\cal E}^{}_{%
{\bf k k^{\prime}}}\! +\omega^{4}\!\int^{}_{{\bf q}}{\cal E}^{}_{{\bf k q}}
D^{}_{{\bf q}}{\cal E}^{}_{{\bf q k^{\prime}}}\!+\!\omega^{6}\!\int^{}_{{\bf %
q}}\!\int^{}_{{\bf q^{\prime}}}{\cal E}^{}_{{\bf k q}} D^{}_{{\bf q}}{\cal E}%
^{}_{{\bf q q^{\prime}}}D^{}_{{\bf q^{\prime}}}{\cal E}^{}_{{\bf %
q^{\prime}k^{\prime}}}\!\ldots\,.
\]
As a result we have 
\begin{equation}
\frac{\partial {\mbox{\large$t$}}^{\omega}_{{\bf k k^{\prime}}}}{\partial%
{\bf k}}\!= \!\omega^2\frac{\partial{\cal E}^{}_{{\bf k k^{\prime}}}}{%
\partial{\bf k}}\!+ \!\omega^2\int\frac{d{\bf q}}{\left(q^{2}\!-\!\omega^{2}%
\right)}\frac{\partial{\cal E}^{}_{{\bf k q}}}{\partial{\bf k}}\,{%
\mbox{\large$t$}}^{\omega}_{{\bf q k^{\prime}}}.
\end{equation}
Here ${\cal E}^{}_{{\bf k k^{\prime}}}\!=\!{\cal E}\!({\bf %
k\!\!-\!\!k^{\prime})}$ is the Fourier transform of the dielectric function
of a scatterer, $\varepsilon({\bf x})$. In the case of a spherical scatterer
with the constant permitivity ${\cal E}({\bf k})$ has the form that is easy
to differentiate ${\cal E}\!({\bf q})\!=\!\left(\sin{qa}\!\!-\!\!qa\cos{qa}%
\right)4\pi\varepsilon /\!{q^3}$.

Eqs.\ (15) and (16) are particular convenient when both momenta ${\bf k}$
and ${\bf k^{\prime}}$ are taken on the mass shell. In this case on-shell
and off-on-shell matrices only appear in the expressions for the derivatives
and one does not have to differentiate the cumbersome Eq.\ (8) and then
calculate the limit of $k\rightarrow\omega$.

\section*{Acknowledgments}

We wish to thank A.Z. Genack for reading and commenting on the manuscript.
This work was supported by the NSF under Grant No. DMR-9311605 and by the
PSC-CUNY research award.


\begin{references}
\bibitem{Kirpatrick}  T.R. Kirpatrick, Phys. Rev. B {\bf 31}, 5746 (1985).

\bibitem{Albada}  M.P. Van Albada, B.A. Tiggelen, A. Lagendijk, and A. Tip,
Phys. Rev. Lett. {\bf 66}, 3132 (1991).

\bibitem{Wigner}  E.P. Wigner, Phys. Rev. {\bf 98},145 (1955).

\bibitem{Boer}  J.F. de Boer, M.P. van Albada, and A. Lagendijk, Phys. Rev.
B {\bf 45}, 658 (1992).

\bibitem{Bart}  B.A. van Tiggelen, A. Lagendijk, M.P. van Albada, and A.Tip,
Phys. Rev. B {\bf 45}, 12233 (1992).

\bibitem{Kogan}  E. Koganand M. Kaveh, Phys. Rev. B {\bf 46}, 10636 (1992).

\bibitem{Cwilich}  G. Cwilich and Y. Fu, Phys. Rev. B {\bf 46}, 12015 (1992).

\bibitem{Kroha}  J. Kroha, C.M. Soukoulis and P. Wolffe, Phys. Rev. B{\bf 47}%
, 9208 (1992).

\bibitem{Soukoulis}  C.M. Soukoulis, S. Datta, and E.N. Economou, Phys. Rev.
B{\bf 49}, 3800 (1993).

\bibitem{Liv1}  D. Livdan and A.A. Lisyansky, J. Opt. Soc. Am. A{\bf 13},
No. 4 (1996).

\bibitem{Liv2}  D. Livdan and A.A. Lisyansky, Phys. Rev. B{\bf 53}, No.
21/22 (1996).

\bibitem{Ozrin}  Yu.N. Barabanenkov and V.D. Ozrin, Phys. Rev. Lett.{\bf 69}%
, 1364 (1992).

\bibitem{Comment}  B.A. van Tiggelen, A. Lagendijk, and A. Tip, Phys. Rev.
Lett. {\bf 71}, 1284 (1993).

\bibitem{Rigorous}  B.A. van Tiggelen and A. Lagendijk, Europhys. Lett.{\bf %
23}, 311 (1993).

\bibitem{Sheng}  {\em The Scattering and Localization of Classical Waves,\/}
ed. by P. Sheng, (World Scientific Press, Singapore, 1990).

\bibitem{Liv95}  D. Livdan and A.A. Lisyansky, J. Opt. Soc. Am. B{\bf 12},
2426 (1995).

\bibitem{Genack}  A.Z. Genack, J.H. Li, N. Garcia and A.A. Lisyansky, in 
{\em Photonic Band Gaps and Localization\/}, ed. C.M. Soukoulis (Plenum
Press, New York, 1993), pp.23-55.

\bibitem{Morse}  P.M. Morse and H. Feshbach, {\em Methods of Theoretical
Physics\/} (McGraw-Hill, New York, 1953).

\bibitem{Abramowitz}  {\em Handbook of Mathimatical Functions\/} ed. by M.
Abramovitz and I.A. Stegun, (Dover, New York, 1972).

\bibitem{Van de Hulst}  H.C. van de Hulst, {\em Light Scattering by Small
Particles\/} (Willey, New York, 1957).
\end{references}
\end{document}